# *Operando* observation of reversible oxygen migration and phase transitions in ferroelectric devices

**Authors:** Pavan Nukala[1,2*], Majid Ahmadi[1], Yingfen Wei[1], Sytze de Graaf[1], Sylvia Matzen[4], Henny W. Zandbergen[3], Bart Kooi[1], Beatriz Noheda[1*]

**Affiliations:**

[1]Zernike Institute of Advanced Materials, University of Groningen, Groningen, 9747 AG, The Netherlands,

[2]Center for Nanoscience and Engineering, Indian Institute of Science, Bengaluru, 560012, India

[3] Kavli Institute of Nanoscience, Faculty of Applied Sciences, Delft University of Technology, 2628 CJDelft, The Netherlands

[4] Center for Nanoscience and Nanotechnology, Paris-Saclay University, CNRS, 91120, Palaiseau, France

*Correspondence to: pnukala@iisc.ac.in, b.noheda@rug.nl

**Abstract:**

Unconventional ferroelectricity, robust at reduced nanoscale sizes, exhibited by hafnia-based thin-films presents tremendous opportunities in nanoelectronics. However, the exact nature of polarization switching remains controversial. Here, we investigate epitaxial $Hf_{0.5}Zr_{0.5}O_2$ (HZO) capacitors, interfaced with oxygen conducting metals ($La_{0.67}Sr_{0.33}MnO_3$, LSMO) as electrodes, using atomic resolution electron microscopy while *in situ* electrical biasing. By direct oxygen imaging, we observe reversible oxygen vacancy migration from the bottom to the top electrode through HZO and reveal associated reversible structural phase transitions in the epitaxial LSMO and HZO layers. We follow the phase transition pathways at the atomic scale and identify that these mechanisms are at play both in tunnel junctions and ferroelectric capacitors switched with sub-millisecond pulses. Our results unmistakably demonstrate that oxygen voltammetry and polarization switching are intertwined in these materials.

**Main Text:**

The discovery of silicon compatible nano-ferroelectricity in hafnia-based thin-films (*1*) has triggered vast amount of research at a fundamental level and rejuvenated interest in ferroelectric materials in microelectronics for low-power non-volatile memory and logic devices. This ferroelectricity is robust even at film thicknesses as low as 1 nm (*2–4*), a situation that was believed impossible according to the classical understanding of ferroelectricity. The spontaneous polarization observed in these films is ascribed generally to a metastable polar orthorhombic phase (*Pca2₁*, o-phase) (*5*). A higher energy rhombohedral phase (*R3m/R3,* r-phases) has been reported for epitaxial growth of $Hf_{0.5}Zr_{0.5}O_2$ (HZO) on $La_{0.67}Sr_{0.33}MnO_3$ (LSMO) buffered perovskite and on trigonal substrates (*6*), and on the former, remnant polarization ($P_r$) values as high as 35 μC/cm² were measured (*7*). The pronounced effects of particle size reduction, surface effects, dopants, oxygen vacancies ($\ddot{V}_o$), epitaxial strain and residual stresses at nanoscale have all been investigated as possible reasons to stabilize these otherwise metastable phases in thin films (*5*, *8*, *9*). On the other end, films as thick as 1 μm also were shown to be ferroelectric, with dopant and defect chemistry being the reason for its stabilization (*10*, *11*).

Armed with this understanding of the virgin state polarization, vibrant research is being conducted on the mechanism of polarization switching. The dynamics of this process through the lens of nucleation limited switching model, pointing out the negligible role of domain growth (or domain wall motion), have been studied by several authors for doped hafnia films grown in the o-phase (*12–18*). Recently, flat phonon bands, and localized dipoles in half unit cells in the o-phase have been postulated as an intrinsic reason for switching without forming domain walls in these systems (*19*).

On the other hand, hafnia-based materials are an important class of resistive memory devices that are known to exhibit memristive hysteresis driven by $\ddot{V}_o$ conduction and redox reactions (*20*, *21*). More generally, in devices of thin-film ferroelectric oxides such as tunnel junctions, both $\ddot{V}_o$ migration and polarization switching lead to memristive hysteresis (*22–27*). Understanding whether these effects are synergetic or independent is crucial to achieve device control, and is a topic of active research (*29–33*). In tunnel junctions of HZO thin-films on LSMO buffered SrTiO$_3$ (STO), Wei et al. (*22*) observed a divergence of the tunnel electro-resistance (TER) from 100% to $10^6$%, upon device cycling, which was explained as a possible transition from polarization switching to $\ddot{V}_o$ migration assisted switching, suggesting the independence of the two mechanisms. Sulzbach et al. (*23*) also reported a similar divergence in the TER as a function of the applied voltage in HZO layers before breakdown. However, it has also been proposed theoretically that electric polarization itself in hafnia originates from oxygen vacancies (*via* electrostrictive effects), strongly suggesting the extrinsic nature of the polarization switching (*29*). Direct structural observations during polarization switching can potentially resolve these controversies (*33*).

Here we report *operando* atomic scale electron microscopy investigations of the behavior of LSMO/HZO/LSMO capacitor stacks grown on conducting (Nb-doped) STO substrate under electric field (see materials and methods in SI). LSMO is a standard choice of bottom electrode in complex oxide devices (*23–26,31*) and, thus, the findings reported here are relevant for understanding a wider class of devices. *In situ* biasing measurements were performed while employing at the same time two scanning transmission electron microscopy (STEM) imaging modes: high-angle annular dark-field (HAADF) STEM and integrated differential phase contrast (iDPC) STEM. With the latter technique we recently succeeded to image hydrogen atoms (lightest element) next to metal atoms (Ti) (*34*), demonstrating that this is currently the most robust atomic resolution imaging technique to measure simultaneously heavy and light elements. Here, by directly imaging oxygen, we provide evidence of the reversible and hysteretic migration of $\ddot{V}_o$ from the bottom to the top electrode through the HZO layer. Associated with such migration, we show $\ddot{V}_o$ induced phase transitions in LSMO (bottom electrode) and HZO layers. Similar mechanisms are also found at short timescales both in cycled tunnel junction (ultra-thin) devices and in ferroelectric capacitors, clearly showing that polarization switching and oxygen voltammetry are not independent mechanisms.

First, we present the evolution of the epitaxial LSMO layer (bottom electrode) with bias, with the voltage applied to the top electrode, keeping the bottom electrode at 0 V (Fig. S1, materials and methods in SI). The virgin state iDPC-STEM image in Fig. 1a. (see also Fig. S2a) shows the antiphase octahedral δ tilts present in the LSMO perovskite structure (see S1 in SI). Mn-O-Mn bond angles, as measured in various such regions, are between 165°-176° (*35*). Upon increasing the bias to 2 V, a very noticeable feature appearing throughout the film (barring the first three monolayers at the interface with Nb:STO) is the displacement of many Mn columns away from the center of oxygen octahedra (Fig. 1b, Fig S2). These deviations indicate a transformation from an $MnO_6$ octahedral towards an $MnO_5$ square pyramidal coordination (see S2 in SI). Thus, at 2 V the LSMO film contains a combination of $MnO_5$ and $MnO_6$ polyhedra. While such a structural feature is not previously observed for LSMO, Brownmillerite (BM, oxygen deficient perovskite) phases are reported to exhibit $MnO_5$ square pyramids in the related material, $SrMnO_3$ (*36*). We will refer to this $MnO_5$ - $MnO_6$ combination as a *BM-precursor* phase in the rest of the article. As for the first 3 monolayers mentioned above, a significant feature is the exaggerated antiphase δ tilts, with Mn-O-Mn bond angles of 143-146°, untypical of perovskite structures (Fig. S2 b-d).

Upon increasing the biasing voltage to 4 V, LSMO converts into a well-studied BM phase (*37,38*) except for the first few monolayers near the interface with the substrate, which transform to the BM-precursor phase (Fig. S3a). Transformation from a perovskite to BM phase occurs via $\ddot{V}_o$ ordering in every alternate Mn-O plane along the *c'*-axis (Fig. 1c) transforming the Mn coordination from octahedral or square pyramidal to tetrahedral. Back-to-back $MnO_4$ tetrahedra along [1-10] alternate with $MnO_6$ octahedra along *c'* signifying the BM phase can be clearly seen in Fig. 1c. This is a hysteretic, non-volatile transformation, and LSMO remains in the BM phase even when external bias is removed (Fig. S3, see S3).

The multiple-step transformations (Fig. 1d) from $MnO_6$ octahedra (virgin state) towards square pyramids plus octahedra (2 V), to alternating octahedra and tetrahedra (4 V) also correlate to the variation of the pseudocubic lattice parameter along the field direction (to be called *c'* from here on). The *c'* values at various bias voltages are shown in Fig. 1e for the first 20 monolayers in LSMO starting from the Nb:STO interface. In the virgin state *c'* is measured to be (384 ± 5) pm with ($\ddot{V}_o$) disorder induced expansion in some planes (see S4 in SI). At 2 V, *c'* oscillates at various values between 335 and 425 pm, without any particular superstructure. At 4 V, except for the first few monolayers, *c'* alternates between 375 and 445 pm, doubling the periodicity. Energy dispersive spectroscopy (see materials and methods in SI) reveals a clear gradient of oxygen concentration in the bottom electrode even at a low bias of 1.5 V, compared to the virgin state, with more $\ddot{V}_o$ occurring closer to the Nb:STO interface (Fig. 1g).

Importantly, the BM phase can be reoxygenated when negative voltages are applied to the top electrode. This is shown to take place for biases as low as -1 V (Fig. S4a). Fig. 2a and b compare iDPC-STEM images of the same field of view at 0 V and -1.3 V. The hysteretic BM phase, clearly begins to reoxygenate as evidenced by the appearance of extra oxygen columns at -1.3 V (Fig 2b) in the Mn-O planes that were oxygen deficient at 0 V (Fig. 2a, Fig. S4a). Upon ramping the bias to -3 V, the entire layer converts to the BM-precursor phase (Fig. 2c) and is retained so when the bias is removed (Fig. 2d), as confirmed by the corresponding 'disorderly' *c'* variation (Fig. S4b).

In order to address the timescales of the processes associated with de- and re-oxygenation of LSMO layers, we followed the dynamics through HAADF-STEM image acquisition, after poling at -4 V (transforming LSMO completely back to the starting perovskite phase). Fig. 2e shows the HAADF-STEM image evolution within 2 min of one particular region, upon ramping the bias from 0 V to 3 V. The initial perovskite phase changes to the BM-precursor phase in 60 seconds, and then to the BM phase within 120 seconds, as indicated by the observed variations in the $c'$ parameter. Then we applied -3 V and observed a complete transformation from a BM phase back to perovskite phase (in this region) within 90 seconds (Fig. 2f, FFT in Fig. S4c, Video S1). Between 60 and 70 seconds we followed the same region by faster HAADF-STEM image series (1.2 seconds/frame). From the $c'$ parameter variations, it can be concluded that the BM-precursor phase transformed to the perovskite phase (Fig. 2f, center panel). Thus, while the complete transformation from perovskite to BM and back takes about a couple of minutes at 3 and -3 V, $\ddot{V}_o$ migration and partial phase-transitions already start occurring in time scales of seconds at these voltages. At 2 V, however, the partial transition to the BM-precursor phase itself takes 3-4 hours. Such speeding of kinetics with voltage is consistent with the ultra non-linear "voltage-time dilemma" typically observed in oxide resistive memories (*39*). Thus, it can be expected that at higher voltages these mechanisms will occur at exponentially smaller time scales.

Next, we describe the structural evolution in the HZO (6 nm) layer under the application of bias. From the multislice iDPC-STEM image simulations for HZO (see also *Ref.* (*40*)) in the r-phase (*R3m*) with [111] out-of-plane (inset of Fig. 3a), we recognize the (001) planes (at ~55º with respect to the [111] direction) by cationic (Hf/Zr) columns surrounded by two oxygen columns on either side of them. In the virgin state, our experimental images perfectly match the r-phase simulations. We followed the evolution of a supercell (Fig. 3a) in this grain upon application of bias along the out of plane [111] direction. The displacement of $\ddot{V}_o$ for this supercell (details in materials and methods, Fig S5a) with respect to the 0 V configuration is shown in Fig. 3b. While $\ddot{V}_o$ migrate towards the bottom electrode with increasing bias (Fig. 3b), we also note that they gather some in-plane displacement (Fig 3b, inset).

At 4 V, the same grain transforms into a combination of multiple grains (Fig. 3c). Upon inspecting various regions in the film, we found that the majority of the grains have changed their structure from r-phase towards orthorhombic (o-) and monoclinic (m-) phases, the thermodynamically more stable phases (Fig. 3d, also Fig. S5c). The o-phase is commonly observed in ferroelectric HZO layers grown by various methods (*5*), and the r-phase is only observed under specific growth procedures and conditions (*6*). These observations on HZO point unambiguously to the fact that the r-phase is stabilized under slight oxygen deficient conditions. Replenishment of oxygen in the HZO layer under bias (originating from the bottom LSMO layer), transforms it into more stoichiometric m- or o-phases. The $\ddot{V}_o$ in the HZO layer, and thus the r-phase, is restored (by reverse migration) upon applying a bias of -3 V, as can be seen from the perfect match of the experimental iDPC-STEM of two representative domains (180º rotated from each other) shown in Fig. 3e, with the multislice image simulations (Fig. 3a-inset). For the devices in the virgin state, and those that underwent full cycle of switching, by assuming a Born effective charge of +4 on cations, we estimate an intrinsic $P_r < 9$ µC/cm$^2$ (see *S5* in SI for $P_r$ estimation and Fig S5b), quite small compared to 35 µC/cm$^2$ observed experimentally (*7*). This discrepancy is a clear evidence that most of the switching charge is related to the oxygen voltammetry.

Ferroelectric devices are normally operated at sub-millisecond timescales, not accessible through *operando* imaging experiments. To understand oxygen migration at these timescales, we have investigated structure-property correlations in different devices with various top-electrodes by carrying out electrical writing and measurements, *ex situ*. First, we present results on the tunnel junction devices Co/HZO (2 nm)/LSMO//STO that were switched (<1 ms pulses) between the low-resistance state (LRS, -6 V) and the high-resistance state (HRS, +6 V), and exhibit an increasing TER from 100% (stage A) to $10^6$% (stage B) upon cycling ~100 times (*22*, *41*) (see Materials and methods in SI). An iDPC-STEM image from a selected region in the LSMO (bottom electrode) layer in the LRS (stage A) (Fig. 4a) shows a clear perovskite structure. In the HRS (stage B), however (Fig. 4b), cobalt oxidizes to $CoO_x$, leaving LSMO oxygen deficient, in the BM-precursor phase.

Next, we report results on ferroelectric capacitor stacks LSMO (or TiN)/HZO (7 nm)/LSMO//STO. From the P-V hysteresis loops at various temperatures (Fig. 4c and inset), we observe that, contrary to what it is expected in classical ferroelectrics, the $P_r$ increases with increasing temperature in the range from 150 K to 300 K (*see S6 in SI*). This is in line with the correlation between polarization switching and oxygen migration, which also improves with temperature.

These devices were prepared in a "down polarized" configuration (5.0 V,1 kHz) and imaged. iDPC-STEM images clearly reveal the oxygen deficient BM-precursor phase in the bottom LSMO layer close to the HZO interface (Fig. 4d) as also confirmed by corresponding disorderly oscillations in the *c'* parameter (inset, Fig. 4d). Close to the STO interface, we find exaggerated oxygen octahedral tilts with Mn-O-Mn bond angles < 146º (marked in Fig. 4e), which is the same interfacial feature observed during *in situ* DC testing at 2 V (Fig. S2b), additionally suggesting that such extreme tilts initiate the transformation from a perovskite to the BM-precursor phase. Furthermore, certain oxygen columns (marked by red circles in Fig. 4e) are visually less intense compared to the rest, consistent with the presence of $\ddot{V}_o$. Thus, the oxygen migration induced structural changes that result from a long time biasing at smaller voltages (hours at 2 V, seconds at 3 V) are the same as those that result after a few cycles of short pulses (sub-milliseconds) at larger bias (~5 to 6 V), in both the ferroelectric capacitors and HZO based tunnel junction devices. Discharge of $\ddot{V}_o$ in the form diffusion-assisted topotactic phase transitions in the electrodes follows the switching (S6 in SI). This process becomes sluggish at lower temperatures, leading to an in-built field which decreases with increasing temperature (Fig. S6).

Perovskite manganites (*42*) are widely used epitaxial bottom electrode layers in various complex oxide-based devices, especially in ferroelectric capacitors and tunnel junctions (*23–26*, *43*). Their crucial role as oxygen conducting memristive layers (*38*) (see *S7* in SI) actively participating in charge transport is being recognized, with recent seminal demonstrations in tunnel junction devices (*27*). Our results show clear and direct evidence that macroscopically measured polarization switching in epitaxial HZO ferroelectric devices is intertwined with oxygen voltammetry (Fig. 4f). This is in line with recent theoretical prediction of electrochemical origin of ferroelectricity in hafnia based thin-films (*29*). We show the exact physical mechanisms taking place during oxygen exchange in ferroelectric HZO devices, with multiple step phase transitions happening concomitantly within the film and the electrode, far beyond their interfaces. Moreover, our work provides deep insights into the reversible transformation of HZO from r-phase to o/m-phases with removal and addition of $\ddot{V}_o$ into the HZO layer, very relevant in the context of doped-hafnia capacitors (initially stabilized in o-phase)*,* interfaced with electrodes having low reduction potential such as Ti, Co, Al, and TiN (*44*).

**Acknowledgments and Funding:** PN would like to acknowledge funding from European Union's Horizon 2020 research and innovation programme under Marie Sklodowska-Curie grant agreement #794954 (nickname: FERHAZ). YW and SM acknowledge a China Scholarship Council grant and a Van Gogh travel grant.

PN acknowledges all the discussions with Pratyush Buragohain from the University of Nebraska.

**Author contributions:** PN, BN and MA conceived the idea. PN synthesized the samples through PLD. MA, PN devised the *in situ* biasing device preparation protocol using FIB, and made the devices. PN, MA and SG set-up the electrical biasing system compatible with Themis-Z microscope. PN, MA and SG carried out the *in situ* biasing experiments with timely help from HWZ and BK. YF and SM synthesized and fabricated tunnel junction and capacitor devices and did the *ex situ* electrical testing. PN, MA prepared the lamellae and carried out imaging experiments. PN, MA, YF analyzed the data. SG carried out image simulations. All the authors discussed the data. PN, BN co-wrote the manuscript, which was read and edited by all the authors.

**Competing interests:** The authors declare no competing interests

**Data and materials availability:** The data is available on University of Groningen data repository. It will be shared with interested parties if the authors are requested.

**Supplementary Materials:**

Materials and Methods

Supplementary text, S1-S7

Fig. S1-S6

**Figure captions:**
**Fig. 1: Deoxygenation of bottom electrode LSMO layer with increasing positive bias.** (a) iDPC-STEM image of a representative region of the bottom LSMO layer in the virgin state, viewed along [110] zone axis, exactly matching the perovskite structure. Schematic in the inset showing $MnO_6$ octhahedra and their antiphase tilts, clearly imaged in (a). (inset) La/Sr: green, Mn: red, O:brown). (b) iDPC-STEM image at V= 2 V. Panels on the right show various unit cells, illustrating Mn columns (circled in white) and their displacements (marked by green arrows) away from the center of an octahedron. Oxygen columns are marked in red circles. (c) iDPC-STEM at 4 V. BM LSMO (zone axis: *a*, schematic in inset) signified by alternating $MnO_4$ tetrahedra and $MnO_6$ octahedra along *c'*. (d) Schematic showing the evolution of an $MnO_6$ octahedra in the virgin state (enclosed in black box) towards $MnO_5$ square pyramids at 2 V (enclosed in red box) to alternating $MnO_4$ tetrahedra and $MnO_6$ octahedra at 4 V (enclosed in blue). (e) Plot of variation of *c'* (La-La distance) parameter from the STO interface in perovskite (black), BM-precursor (red) and BM phases (blue). (f) Overview image of LSMO/HZO/LSMO capacitor with regions marked where oxygen content was quantified from EDS (g) at 0 V and 1.5 V. Scale bars, 1 nm in (a,b, c), and 5 nm in (f).

**Fig. 2: Oxygenation of bottom electrode LSMO layer with increasing negative bias.**
iDPC-STEM images of a region in the same field of view at 0 (a) and -1.3 V (b). At -1.3 V oxygen columns start to appear (marked by red arrows) in positions where there were none in the BM phase at 0 V. (c) BM phase transforms to BM precursor phase at -3 V and is retained so at 0 V (d). (e) Dynamics is recorded via HAADF-STEM imaging within 120 seconds of ramping from 0 to 3 V from a starting perovskite phase. A BM-precursor phase is imaged at 60 sec, BM phase is imaged at 120 sec. (f) Upon changing the bias to -3 V, a BM phase is recorded at 30 sec, changes to perovskite phase by 90 sec (disappearance of the superstructure spots in FFT, Fig. S4). The intermediate BM-precursor phase recorded at 60 sec, converts to perovskite-like phase in about 6 sec as can be seen by the variation in c' parameter, giving an idea about the time scales of $\ddot{V}_o$ migration. Scale bars: 1 nm in (a-d) and 5 nm in (e, f)

**Fig. 3: Oxygenation and deoxygenation of HZO and associated phase transitions.** (a) Evolution of an *r-phase* HZO grain while oxygenating under positive bias followed through iDPC-STEM image, where both cations and oxygen columns are displayed. The image of the virgin state

at 0V shows two red arrows pointing to two oxygen columns in the (100) planes neighboring a cationic column. Multislice iDPC-STEM image simulations of the r-phase (*R3m* symmetry) in the inset shows the good match with the observations. (b) Out-of-plane displacement of $\ddot{V}_o$ with external bias, in the marked supercell (red box) with respect to the positions in (a). Negative values indicate displacement towards bottom electrode. $\ddot{V}_o$ shows both in-plane and out-of-plane (towards bottom electrode) components (inset). (c) A new grain nucleates in the same region at 4 V, giving rise to a polycrystalline nature (FFT in inset). (d) Another region in the HZO film back at 0 V showing o-phase and m-phase (with multislice simulations of both in the insets). Also note the change of orientation from [111] to [100]. (e) iDPC-STEM image of domains (mutually rotated by 180° about [111]) in the r-phase (to be compared with simulation in inset of Fig. 3a, from ref. *(36)*), which is retained when poled at -3 V (imaged at 0 V). Scale bars: 1 nm in (a,c,e) 2 nm in (d). Interfaces between HZO and top and bottom LSMO are marked in orange. In (d) only HZO/bottom electrode interface is shown.

**Fig. 4: Cycled tunnel junctions and ferroelectric capacitors.** (a) Co/HZO (2 nm)/LSMO tunnel junction in LRS (500 μs, 100s of cycles poled at -6 V). HAADF-STEM image on the left (scale bar: 5 nm), and iDPC-STEM of a selected region in LSMO on the right (scale bar: 1 nm). (b) Ex-situ imaging of tunnel junction in HRS (500 μs, 100s of cycles, poled at 6 V). (right) HAADF-STEM (and corresponding information from EDS) shows CoOx/HZO/LSMO stack (scale bar: 5 nm). (left) iDPC-STEM of a region in LSMO shows BM-precursor phase (scale bar: 1 nm). Selected Mn column displacements towards a square pyramidal geometry are marked by green arrows. (c) Temperature dependent P-V loops obtained from dynamic hysteresis measurements at 1 kHz in LSMO/HZO (7 nm)/LSMO (30 nm) capacitors. Inset shows the corresponding $2P_r$ vs T curve (see also Fig. S6). (d) iDPC-STEM image of LSMO close to HZO interface (marked in orange) in TiN/HZO (7 nm)/LSMO capacitors, prepared in "down polarized" state (at 5 V). BM-precursor phase is formed, confirmed by disorder in *c'* parameter (inset). (e) iDPC-STEM image of LSMO close to the substrate interface (marked in orange) with extreme octahedral tilts. Some low-contrast oxygen columns (containing $\ddot{V}_o$) are marked by yellow circles. (f) Schematic of the polarity dependent oxygen voltammetry process in the Metal-Insulator-Metal device structure. Oxygen deficient phases in various layers are represented with lighter colors.

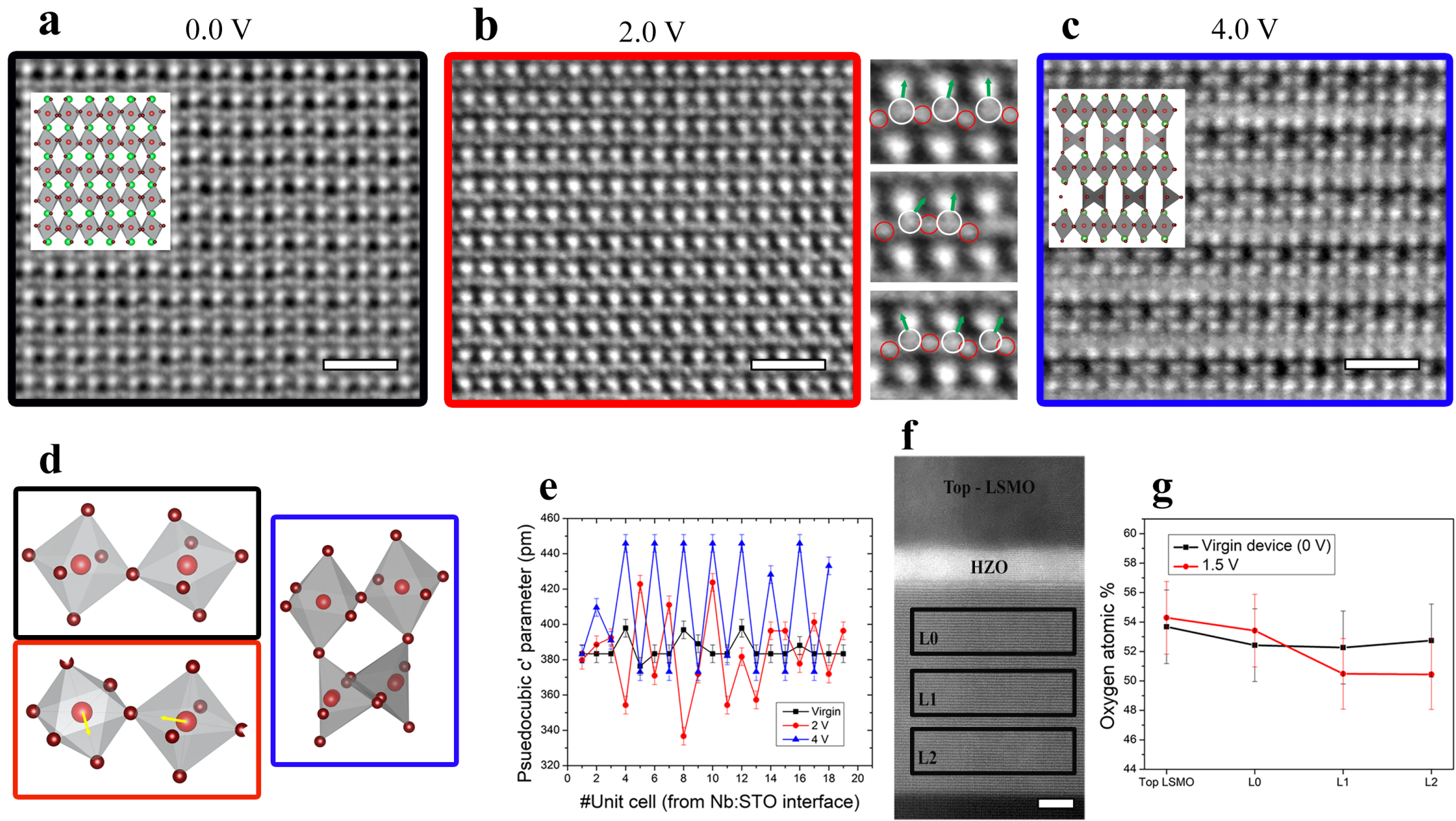

**Figure 1**

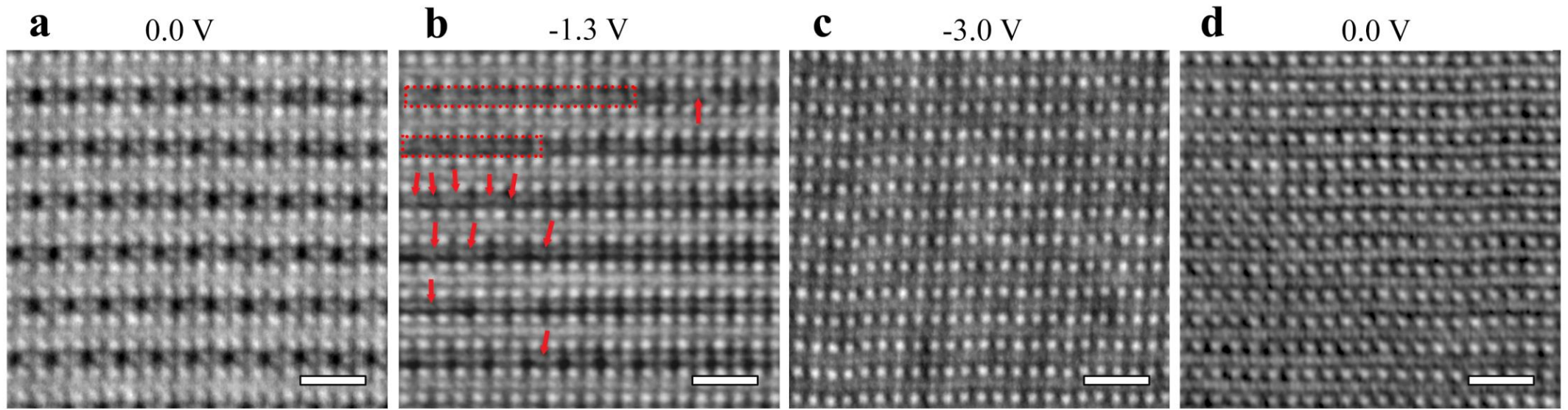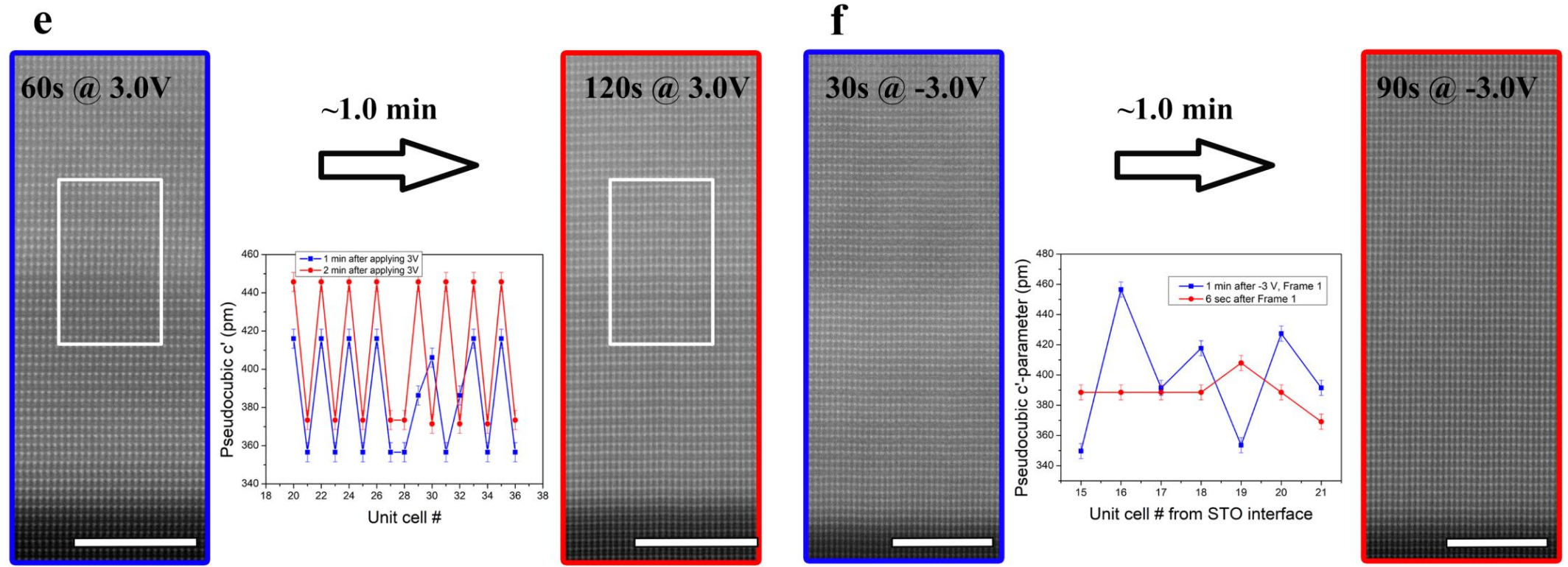

**Figure 2**

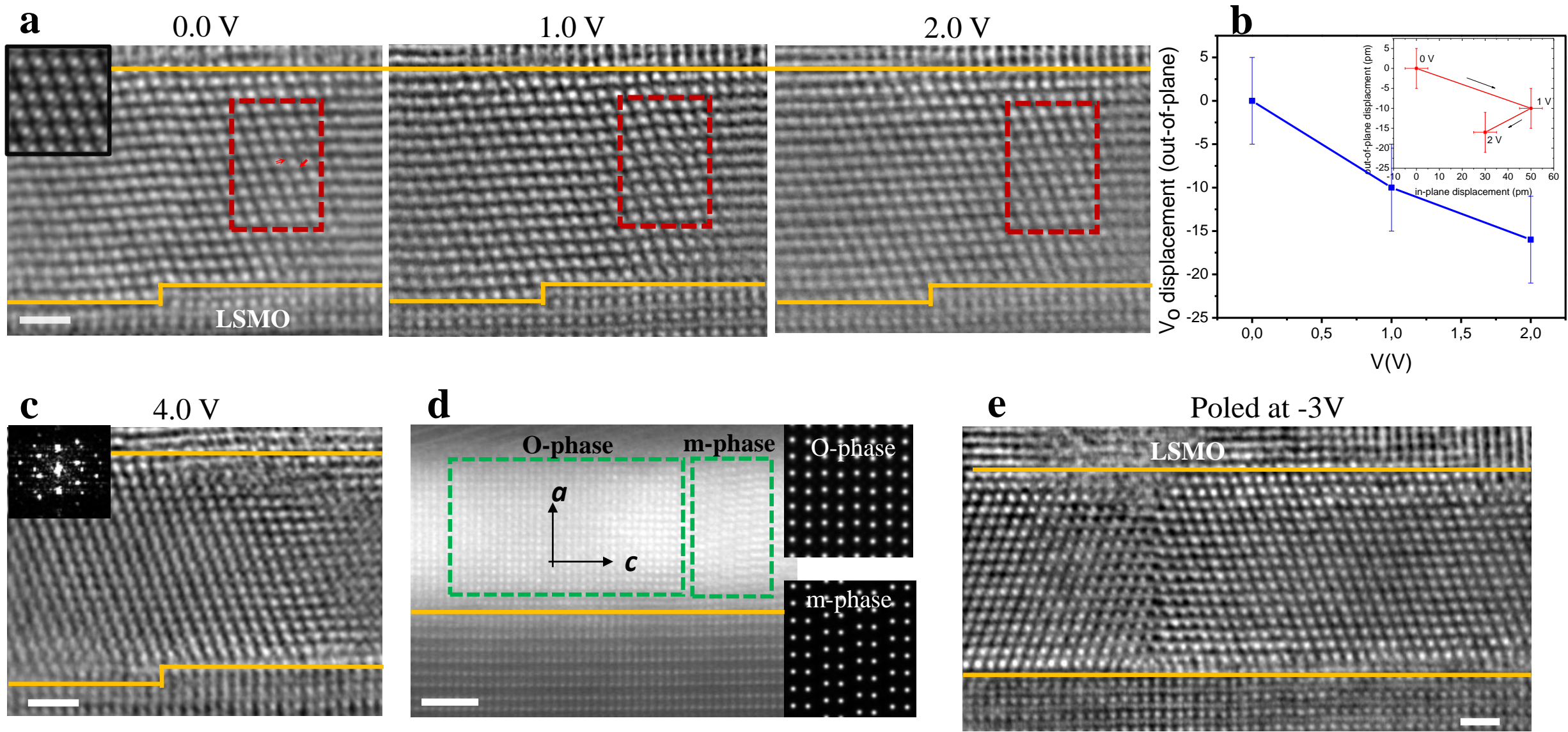

Figure 3

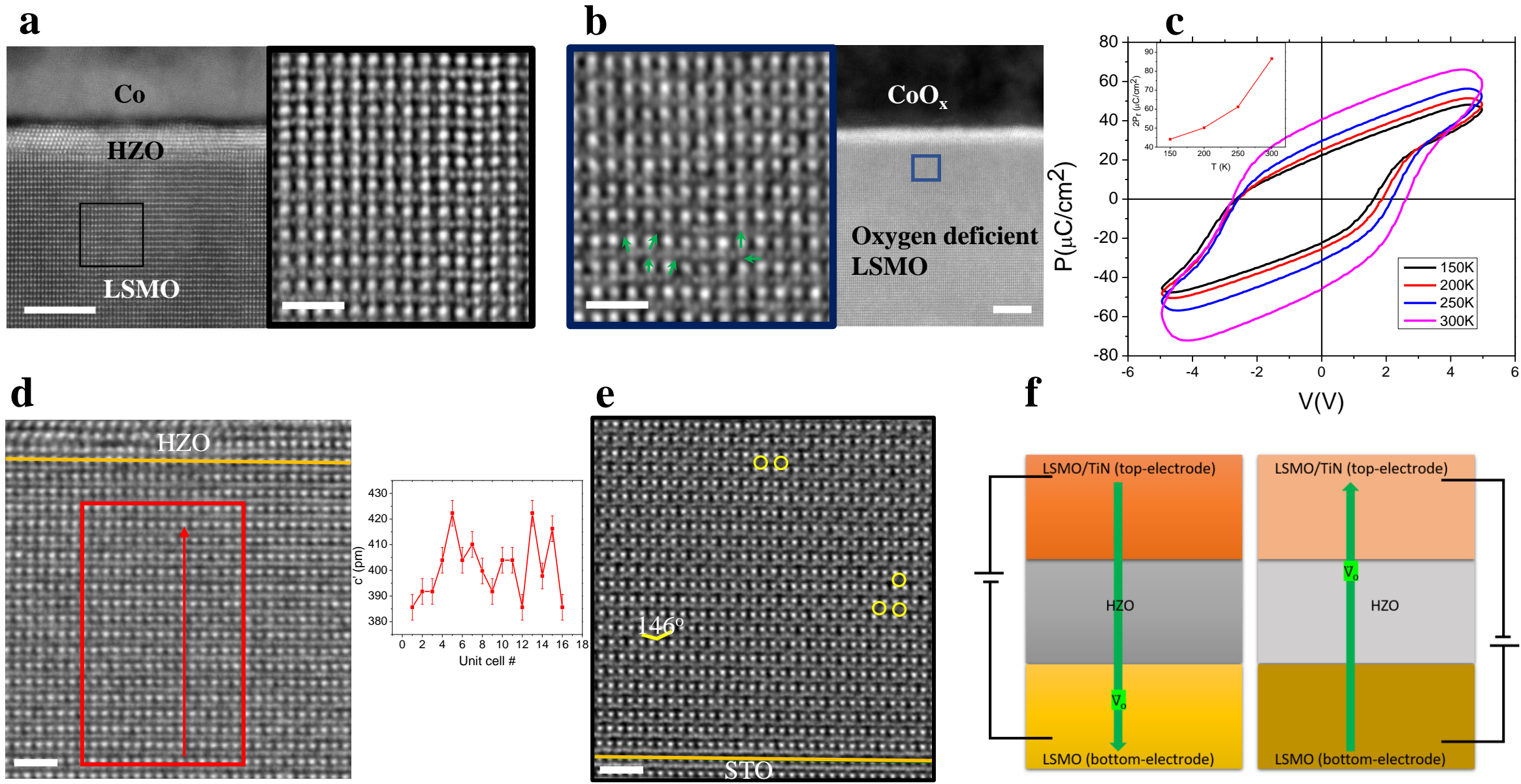

Figure 4

# Supplementary Materials for

## *Operando* observation of reversible oxygen migration and phase transitions in ferroelectric devices

Pavan Nukala[1,2*], Majid Ahmadi[1], Yingfen Wei[1], Sytze de Graaf[1], Sylvia Matzen[4], Henny W. Zandbergen[3], Bart Kooi[1], Beatriz Noheda[1*]

[1] Zernike Institute of Advanced Materials, University of Groningen, Groningen, 9747 AG, The Netherlands

[2] Center for Nanoscience and Engineering, Indian Institute of Science, Bengaluru, 560012, India

[3] Kavli Institute of Nanoscience, Faculty of Applied Sciences, Delft University of Technology, 2628 CJDelft, The Netherlands

[4] Center for Nanoscience and Nanotechnology, Paris-Saclay University, CNRS, 91120, Palaiseau, France

*Correspondence to: pnukala@iisc.ac.in, b.noheda@rug.nl

**This PDF file includes:**

    Materials and Methods
    Supplementary Text, S1-S7
    Figs. S1 to S6

**Materials and Methods**

**Synthesis by Pulsed Laser Deposition:** LSMO (30 nm)/HZO (6 nm)/LSMO (30 nm) layers were grown epitaxially on Nb:STO via pulsed laser deposition (PLD). The growth conditions are described elsewhere (*7*). Pt (200 nm) was sputtered on the top-electrode (LSMO).

*In situ* **TEM set-up, imaging and biasing techniques:** The set-up of the holder, biasing system in Thermo Fisher Scientific Themis Z (monochromated and double corrected) microscope is shown in Fig. S1a. In-situ biasing cum heating customized double tilt holder (HennyZ, Fig. S1b) was used to host the cross-sectional devices (zone axis: [110] Nb:STO) prepared on a MEMS based $SiN_x$ chip (Fig. S1c,d preparation procedure described in the next section). Biasing was performed through Keithley 4200 parameter analyzer by using two SMUs (source measuring units, Fig S1a). Atomic resolution imaging (while biasing) was performed at 300 kV through HAADF-STEM, and more importantly integrated differential phase contrast (iDPC) STEM to image low Z elements such as oxygen. Beam convergence angle was 22.4 m rad. Data from segmented detectors were acquired under the collection angles from 6-23 mrad. H(M)AADF-STEM images were simultaneously acquired with collection angles > 23 mrad. Energy dispersive spectroscopy (EDS) was performed using two large area detectors in total capturing 1.76 steradian. Spectra were acquired from various regions in the capacitor structure at a combined signal of 10-15 kcps for over 1 hour. Voltage is always applied to the top electrode, with bottom electrode at 0 V (Fig S1e).

*In situ* **biasing device fabrication:** MEMS based micro heater & electrical biasing chip (developed by HennyZ, see Fig. S1c,d,e) is used for hosting the TEM cross-sectional lamella, and connecting the lamella to the external pads. This chip has 6 electrodes, 3 on each side of the $SiN_x$ membrane. The outermost electrodes are connected to the microheater and can heat the TEM lamella very locally up to 1000 ºC. The innermost Pt-electrodes are used for biasing. A focused ion beam (FIB, FEI Helios G4 CX) based sample preparation protocol was developed in-house for *in situ* biasing experiments. Electron transparent regions were created in the region between the inner most electrodes of interest on the $SiN_x$ membrane by drilling hole using Ga beam. Cross-sectional chunks (dimensions: 10 x 1.5 x 5 μm$^3$) were initially thinned down to 300-500 nm thickness using standard focus ion beam (FIB) based processing at 30 kV , and then transferred onto the $SiN_x$ membrane of the chip (Fig. S1e). The top (Pt) and bottom electrode (substrate) were then connected to the electrodes on the chip through e-beam induced Pt deposition (EBID). Certain windows (~1 x 3 μm$^2$) on the lamella were thinned down to ~50 nm using Ga beam at 16 kV accelerating voltage, and finally several low kV cleaning steps (5 and 2 kV) were used not only to clean the side surfaces of the lamella, but also to remove the excess Pt deposition (Fig S1e). Electrical measurements were performed *ex situ* after every step to understand any possible leakage sources, and further milling and cleaning steps were performed to remove these sources.

**Regular device fabrication:** Tunnel junction devices (Co/HZO (2 nm)/LSMO) were fabricated using the procedure described in ref. *41*. Two types of ferroelectric capacitors were fabricated: TiN/HZO (7 nm)/LSMO (30 nm)//STO devices were made by depositing TiN as the top electrodes using standard photolithography and sputter deposition followed by lift-off procedure on the rest of the stack grown by PLD. LSMO/HZO/LSMO capacitors were also fabricated. The entire stack was first grown using PLD. Then devices were lithographically defined by etching the top LSMO, HZO and 6 nm of bottom LSMO layer away from the regions of interest using Ar ion-beam etching. Cross sections from these devices were made through standard FIB based TEM cross-sectional lamella preparation procedure.

The tunnel junction devices were tested *ex situ* with +6/-6 V sub-millisecond rectangular pulses for 100s of cycles. Cross-section TEM lamellae were made from one sample in the high resistance state, and another one in the low resistance state. The TiN/HZO/LSMO capacitors were tested for ferroelectric switching using the PUND scheme at +/- 5.5 V using triangular pulses of rise and fall time of 250 μs for tens of cycles. Results presented in Fig. 4c-e are from a cross-sectional lamella of such capacitor, which was eventually ferroelectrically switched "down" with a +5.5 V pulse (Fig. S6).

**Energy dispersive spectroscopy quantification:**

EDS spectra of LSMO recorded with Themis Z microscope were quantified after background subtraction using the Brown-Powell model *(45)*. Oxygen atomic percentages in Fig. 1g were determined from 20 x 5 $nm^2$ regions with one on the top electrode, and three in the bottom electrode with varying distance to the HZO as marked L0, L1 and L2 in Fig. 1f.

**$\ddot{V}_o$ displacement analysis in HZO layer:**

The cationic lattice of the supercell was first constructed by identifying the cation positions through a peak finding algorithm after assigning a finite width defined by a mask around each peak. The rest of the lattice in the supercell is the anionic lattice (see Fig. S4). We quantify the $\ddot{V}_o$ migration as a function of bias in this supercell by estimating the difference between the centers of mass (CoM) of the cations and anions. Under the assumption that cations do not move at small voltages, we can relate this quantity directly to the migration of V̈o. The choice of the mask defining the size of the cations, and the pixel size determine the errors in estimation of the displacement, although peak fitting can occur with sub-pixel size precision.

**P-V hysteresis measurements:**

P-V loops were measured using a dynamic hysteresis measurement scheme on TiN/HZO/LSMO//STO capacitors. A triangular voltage waveform of 1 kHz from 0 to 5 to -5 to 0 V (defining one period) were applied to the device at various temperatures, and corresponding currents were measured. Polarization was calculated using the formula $P = \frac{1}{A} \int i \, dt$, where A is the device cross-sectional area, $i$ is the measured current, and $t$ is the time.

**Supplementary Text**

**S1**. The LSMO layer in the as-deposited film is oriented with $[001]_{pseudocubic}$ (c-axis) out of plane, and observed along the $[110]_{pseudocubic\ (pc)}$ (zone axis). Schematics of A-site (La/Sr), B-site (Mn) and oxygen column arrangements are shown in Fig. 1a. The $a^-a^-a^-$ octahedral tilts typical of rhombohedral LSMO reveal themselves as antiphase δ tilts in this zone (alternating positive and negative Mn-O-Mn bond angles)(*35*). Deviations of Mn columns from octahedral coordination in the virgin state, although present are very minimal, and can be considered to be disorder induced perturbations of the standard structure. Even some disorder in the octahedral tilt angles (165-176°) can be seen in Figs 1a and S2a.

**S2**: Square pyramidal coordination is typical of the BM phases of strontium manganite (SMO, $SrMnO_{2.5}$, Fig. S2d). When BM-SMO is viewed along $[110]_{pc}$ zone axis, it shows two important features (i) Mn displacements away from the octahedral coordination along $[1\text{-}10]_{pc}$ and (ii) a resolvable doubling of alternate oxygen columns between Mn columns with ~90 pm projected distance between the columns along $[1\text{-}10]_{pc}$ (horizontal direction in Fig.1b, Fig. S2b,d). Both these features, characteristic of square pyramidal coordination are present in the BM precursor phase of LSMO (Fig S2c, Mn displacement marked in green arrows, Oxygen column separation marked by blue arrows), albeit in randomly varying directions (not just along [1-10]).

The reported changes in structure at 2 V (Fig. 1b) are hysteretic and remain upon removing the external bias.

**S3**: In the BM phase shown in Fig. 1c (viewed along a-axis), the perovskite-like layer consists of $MnO_6$ octahedra, whereas the oxygen deficient Mn-O layer consists of $MnO_4$ tetrahedra, which along the [1-10] direction alternate between pointing towards each other and away from each other. Also observable in another domain viewed along c-axis (Fig. S3a), where Mn columns in the oxygen deficient Mn-O layer appear more symmetrically arranged. In Fig. S3b, a region is shown with the field of view of both the BM domains.

**S4**: Oxygen vacancies are known to cause some lattice expansion even in virgin samples. These vacancies are typically a result of not just the growth conditions of the film, but also the fact that LSMO grows under tensile strain on STO and Nb:STO (which promotes oxygen vacancies). See ref (*6*).

**S5: The saturation polarization** ($P_s$) of the unit cell (as shown in Fig. S5c) is estimated in 2 different ways.

  (i) Through a peak-fitting algorithm the anionic positions ($y_i$) and cationic positions ($x_i$) were obtained in a unit cell. Then $P_r=\Sigma(4x_i-2y_i)/V$, where V is the Volume of the unit cell (4 and -2 are the Born effective charges of Hf/Zr and oxygen respectively). The unit cell was chosen such that $P_r = 0$ in the symmetrized paraelectric phase (ref. *7*). This yielded $P_r < 2$ µC/cm$^2$ (displacement < 10 pm) for HZO across 14 different unit cells in the *R3m* phase. This procedure assumes HZO is completely stoichiometric, and $P_r$ was similarly estimated on the *R3* phase in ref .*40*. Note that for similar samples in *R3m* phase, $P_r=35$ µC/cm$^2$ was measured through macroscopic electrical switching experiments.

  **(ii)** With the new information from this work on the oxygen deficiency of the *R3m* HZO phase, the stoichiometric assumption presented in (i) does not hold. So, we calculated the $P_r$ by estimating the center of mass of the cationic lattice and the non-cationic part (anionic lattice) in every unit cell (Fig. S5c). The difference in the center of masses gives a displacement of ~38 (±8) pm in the unit cell shown in Fig. S5c (from the domain on the left in Fig. 3e in the manuscript). This suggests that $\ddot{V}_o$ is responsible for almost 4 times increase in the displacement (compared to the stoichiometric lattice in (i)). Whether this will cause significant improvement in direct estimation of $P_r$ cannot be established without the knowledge on the Born charges of cations and anions in this oxygen deficient HZO lattice. Upon assuming that the cationic born charges in this configuration still remains +4, we estimate $P_r < 9$ µC/cm$^2$. In any case, there exists a clear mismatch with the value obtained from electrical measurements, strongly supporting that most of the switching arises from oxygen voltammetry process.

**S6:** The increase of remnant polarization ($P_r$) with temperature presented in Fig. 4c (see also Fig. S6) is consistent with the fact that $\ddot{V}_o$ migration kinetics improves with temperature, clarifying that polarization switching and $\ddot{V}_o$ are intertwined mechanisms. In conventional ferroelectrics, one expects a degradation of $P_r$ with increasing temperature owing to increased thermal fluctuations. Note that similar unconventional behavior was measured by Fengler and coworkers from 25°C to 125°C on ALD deposited HZO samples with TiN electrodes in ref *46*.

Another interesting observation from these measurements is the pronounced presence of imprint (internal field) at low temperatures, which decreases towards zero at room temperature. This can be rationalized as a result of increased sluggishness of the oxygen diffusion during the topotactic phase transitions that follow oxygen migration (discharge mechanism) at low temperatures, resulting in accumulation of charged vacancies and thus space charge (*32*). However, detailed modeling of this interesting aspect belongs to the gamut of further work.

**S7:** Another implication of the present work is the estimation of the field drop across the insulator in Metal-Insulator-Metal stacks. With an oxygen conducting memristor such as LSMO as one of the metals, (e.g. Metal-HZO-LSMO) stacks, field drops (in context of coercive field estimations) across the insulator are typically overestimated due to not considering the field drop in the LSMO layer when oxygen begins to migrate.

**Supplementary Figures**

**Fig. S1**

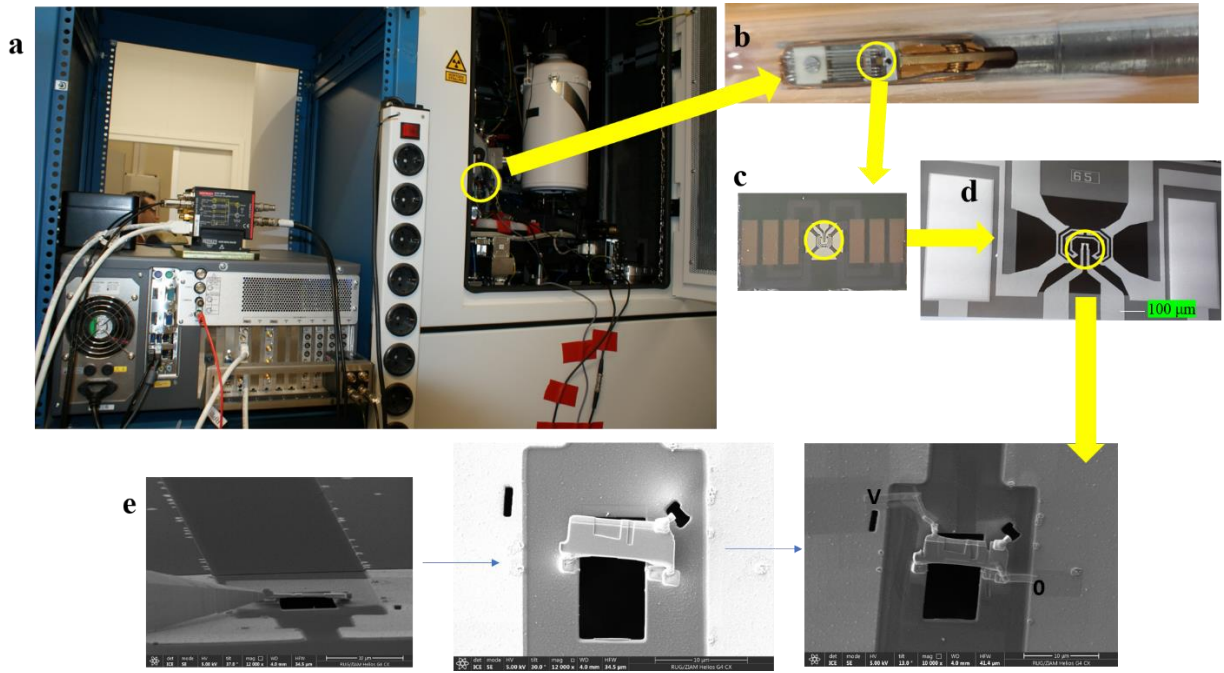

**In-situ biasing set-up and sample preparation**

(a) Set-up showing Keithley 4200 parameter analyzer with two SMUs connected through a switch box (PMU) to the two leads of an in-situ biasing holder, inserted inside the electron beam column. (b) Optical photo of the biasing/heating holder with the inner two leads for biasing, and outer four to resistively heat the MEMS chip, shown in (c). (d) SEM image of the $SiN_x$ (300 nm) membrane hosting the micro-heater, as well as biasing electrodes (inner most). (e) FIB based lamella transfer procedure onto the $SiN_x$ membrane with electron transparent hole created *a-priori* (protocol described in materials and methods).

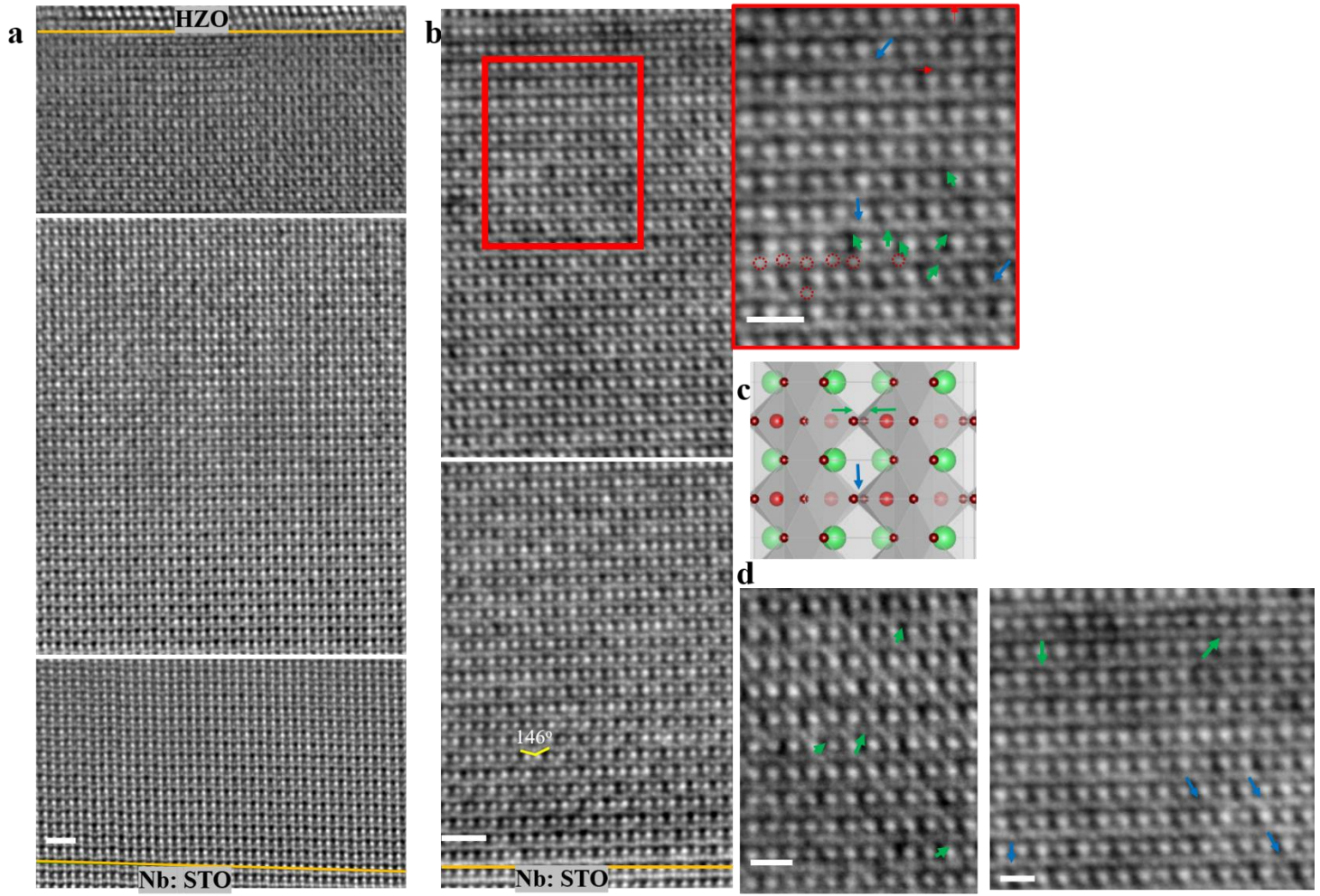

**Understanding the structural features of perovskite and BM-precursor phases**

(a) Overview iDPC-STEM stitched images in the virgin state. (b) Overview iDPC-STEM stitched images at 2 V. Region enclosed in the red box is zoomed in the inset. Exaggerated octahedral tilts (143-146°) can be seen close to the interface with Nb:STO. In the inset of b, apart from marking selected oxygen column positions (red dotted circles), and Mn displacements from the center of an octahedron (green arrows), we also point out to some of the resolvable double oxygen columns between Mn columns (blue arrows). This is another feature of transformation towards square pyramidal coordination of Mn, as illustrated from the schematic of BM $SrMnO_{2.5}$ (SMO) in (c). (d) iDPC-STEM images from other regions (also in BM-precursor phase), where also Mn displacements (green arrows), and resolvable separation between oxygen columns between Mn columns (blue arrows) are marked. Scale bars, 1 nm in (a, b), 500 pm in (d) and inset (right panel) of b.

**Fig. S3**

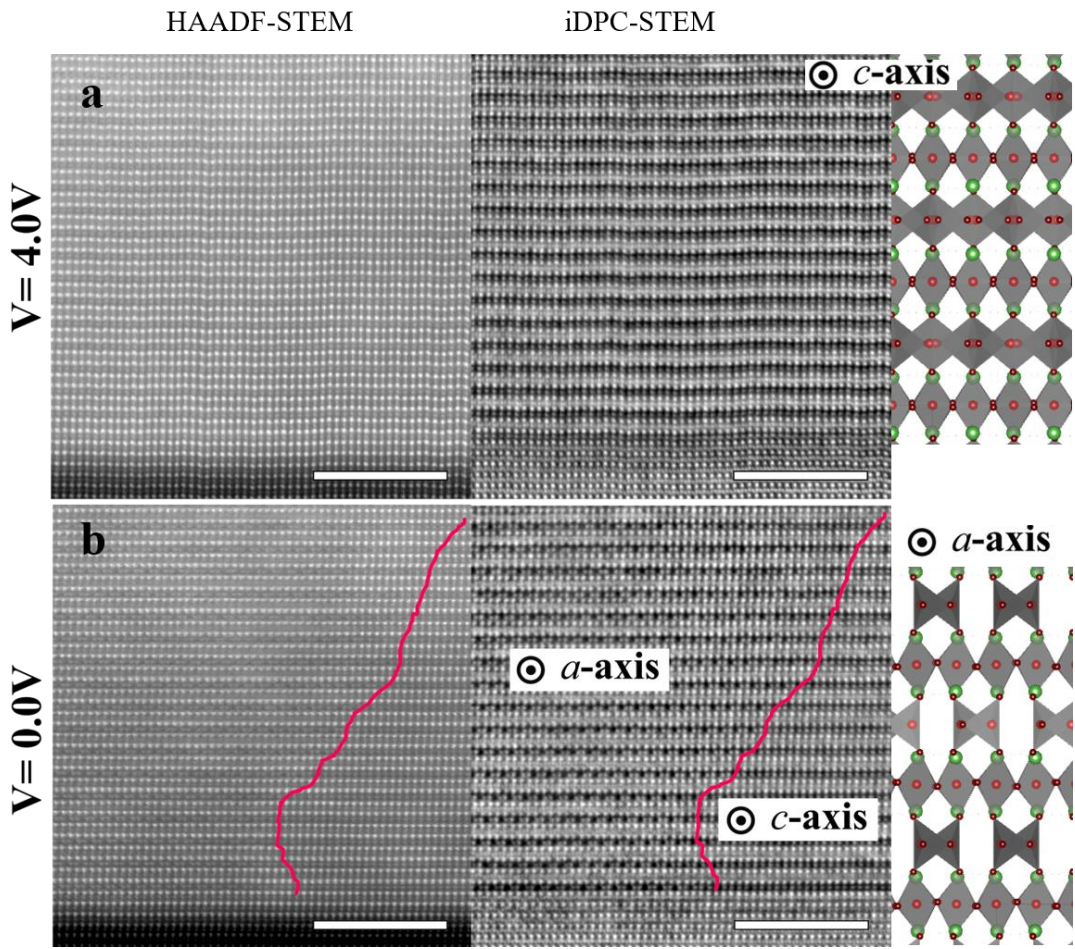

**Hysteretic BM phase:** (a) HAADF-STEM and corresponding iDPC-STEM images of BM phase formed at 4 V. This is a domain whose c-axis is parallel to the viewing direction (zone axis). Schematic of this domain is also shown on the right most panel (La/Sr: green, Mn: red, O: brown). By comparing HAADF and iDPC-STEM images, it can be clearly noted that the first 2-3 monolayers of LSMO near the interface with the substrate are still in a BM-precursor phase. (b) HAADF-STEM and corresponding iDPC-STEM images of BM phase hysterically retained at 0 V. In the field of view, two domains of the BM phase can be noted (beyond the first 3 monolayers of LSMO near the substrate). They have $a$ and $c$ as the zone axes respectively. The back-to-back tetrahedra in alternate Mn-O planes along [1-10] direction can be clearly observed from the domain viewed along a-axis (schematic in the right panel).

Fig S4

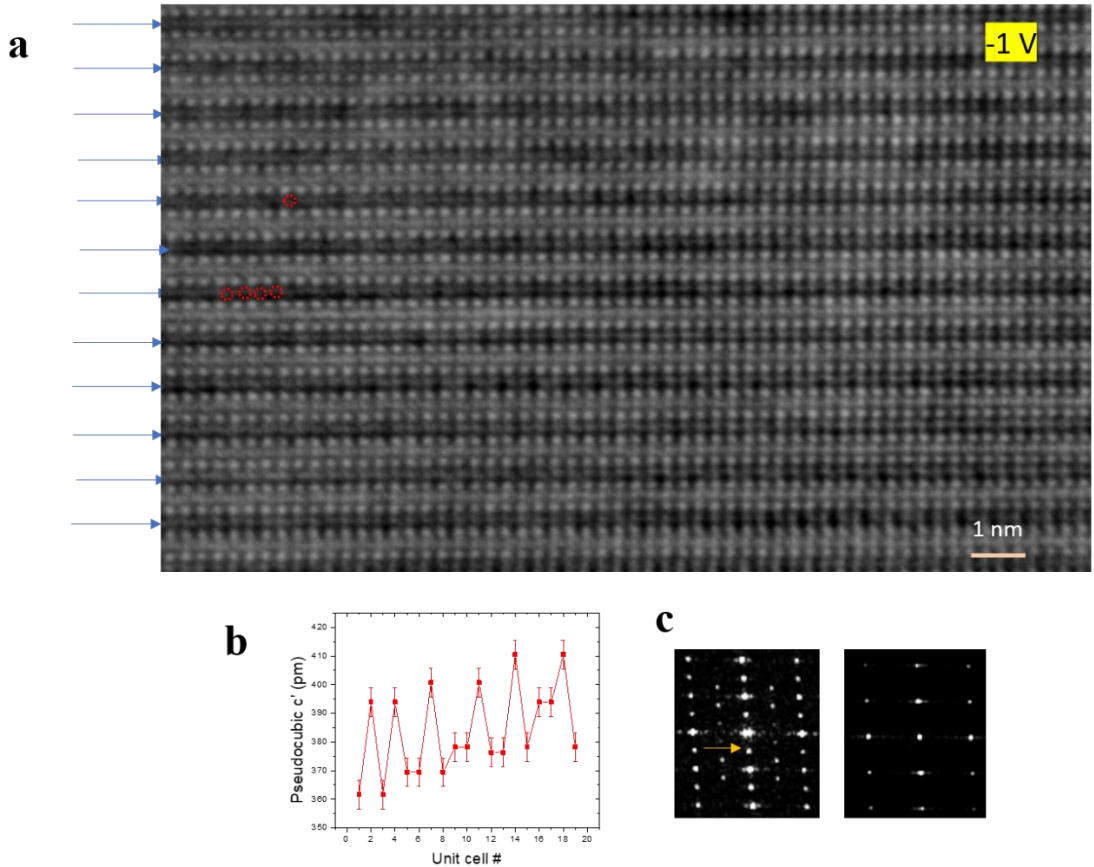

**Reoxygenation of LSMO at negative bias and dynamics**

(a) This iDPC-STEM image is obtained from a representative region in LSMO which has initially transformed to a BM phase upon poling at 4 V. Then, by applying -1 V, we clearly see oxygen columns appearing (some of them marked by red circles as a guide to eye) in the Mn-O planes in darker contrast (marked by blue arrows on the left). The appearance of these new positions also marks a transformation of this Mn-O layer with Mn in tetrahedral coordination to a Mn in higher coordination. (b) $c'$ parameter oscillation across 20 unit cells of LSMO corresponding to Fig. 2c (poled at -3 V), reminiscent of the BM precursor phase. (c) Evolution of BM phase to perovskite phase within 90 seconds of changing the voltage from -3 V to +3 V, followed through Fast Fourier

Transforms (for images presented in Fig. 2e). Notice the disappearance of the superstructure reflections upon this transformation.

Fig. S5

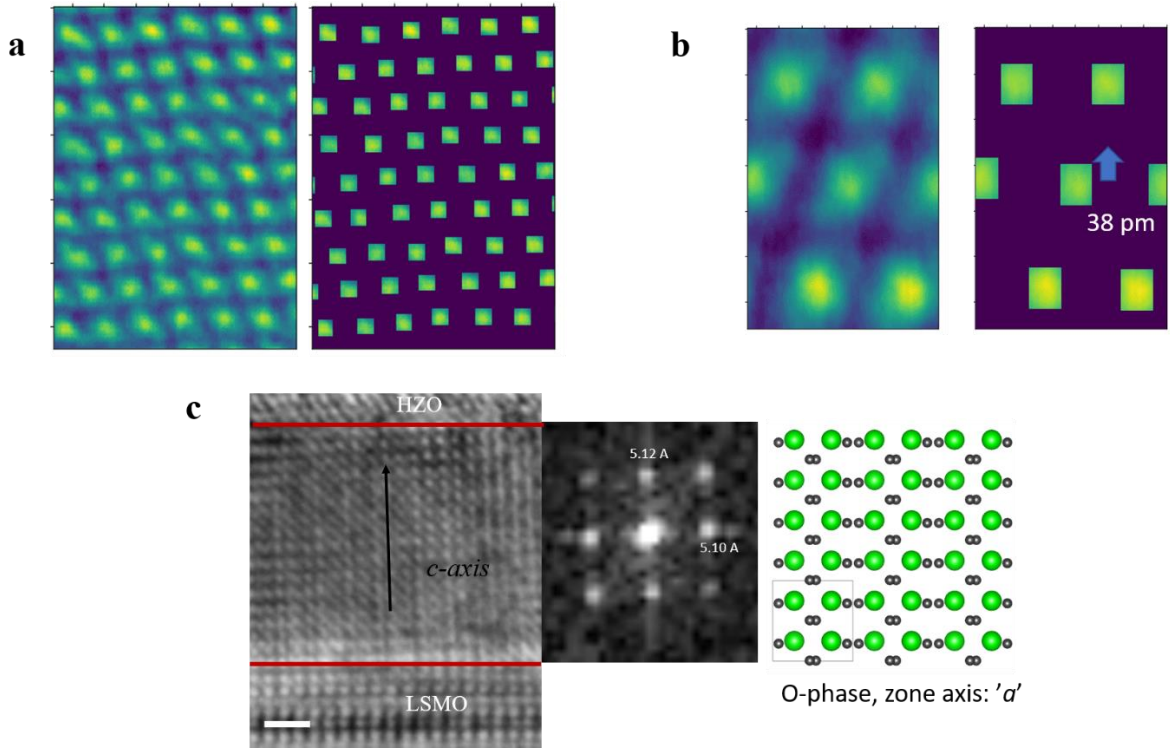

**Evolution of HZO layer**

(a) The iDPC-STEM image of supercell of HZO r-phase grain presented in Fig. 3a at 1 V (left). Cationic lattice is shown on the right. The rest (difference image) is the anionic lattice. Difference in the CoM of cationic and anionic lattice reveals the $\ddot{V}_o$ displacement as a function of bias (Fig. 3b). (b) *R3m* unit cell (iDPC-STEM image of the left domain in Fig. 3e) considered for direct $P_r$ estimations. Upon symmetrizing this unit cell, $P_r=0$ in the paraelectric phase, and this justifies the selection of the unit cell. CoM cation−CoM non cationic part gives the displacement (38 pm in the unit cell shown here). This procedure includes the presence of $\ddot{V}_o$, information present in the raw image contrast. (c) iDPC-STEM image of an o-phase with c-axis out-of-plane, observed at 4 V. FFT and the matching of experimental image to the real space schematic with Hf/Zr columns in green, and oxygen in grey confirm an o-phase.

**Fig. S6.**

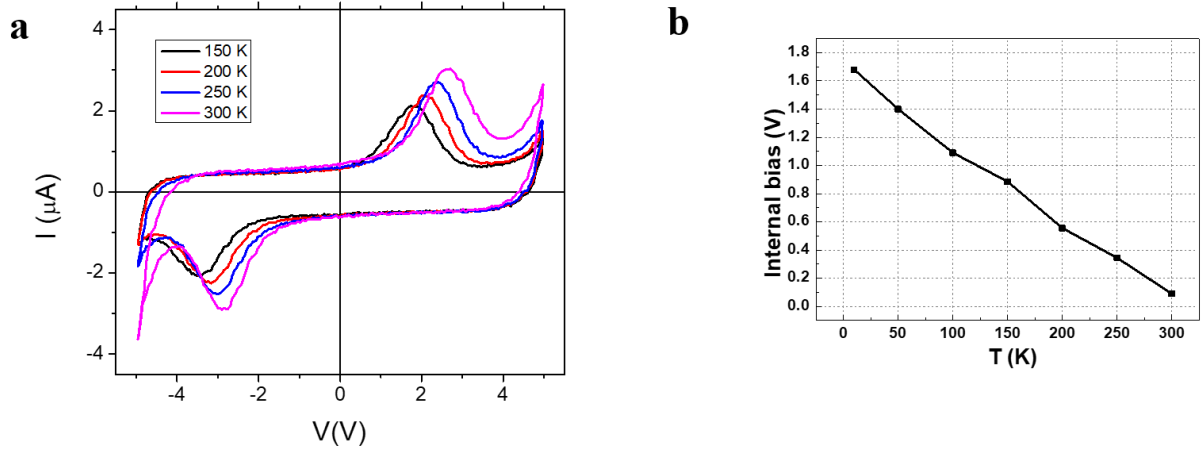

**I-V loops and evolution of imprint on TiN/HZO/LSMO capacitors with temperature**

(a) I-V loops corresponding to the P-V loops presented in Fig. 4c of the manuscript. The $2P_r$ increases with increasing temperature, as shown in the inset of Fig. 4c. (b) Internal bias, calculated as $V_{c+}+V_{c-}$ as a function of temperature, where $V_{c+}$ ($V_{c-}$) are coercive voltages under positive (negative) bias.